\documentclass[11pt, executivepaper]{article}
\usepackage[utf8]{inputenc}
\usepackage[T1]{fontenc}
\usepackage{natbib}
\usepackage{amsmath}
\usepackage{xcolor}
\usepackage{amsfonts}
\usepackage{graphicx}
\usepackage{enumitem}
\usepackage{geometry}
\usepackage{hyperref}
\hypersetup{colorlinks= true, allcolors=blue}
\setcitestyle{aysep={}}

\begin{document}

\title{\textbf{Classical Logic in Quantum Context}}

\author{Andrea Oldofredi\thanks{Contact Information: Universit\'e de Lausanne, Section de Philosophie, 1015 Lausanne, Switzerland. E-mail: Andrea.Oldofredi@unil.ch}}

\maketitle

\begin{abstract}
It is generally accepted that quantum mechanics entails a revision of the classical propositional calculus as a consequence of its physical content. However, the universal claim according to which a new quantum logic is indispensable in order to model the propositions of \emph{every} quantum theory is challenged. In the present essay we critically discuss this claim by showing that classical logic can be rehabilitated in a quantum context by taking into account Bohmian mechanics. It will be argued, indeed, that such a theoretical framework provides the necessary conceptual tools to reintroduce a classical logic of experimental propositions in virtue of its clear metaphysical picture and its theory of measurement. More precisely, it will be showed that the rehabilitation of a classical propositional calculus is a consequence of the primitive ontology of the theory, a fact which is not yet sufficiently recognized in the literature concerning Bohmian mechanics. This work aims to fill this gap.
\vspace{4mm}

\end{abstract}
\vspace{5mm}

\begin{center}
\emph{Accepted for publication in Quantum Reports}
\end{center}
\clearpage

\tableofcontents

\section{Introduction: The Relation Between Logic and Mechanics}
\label{intro}

The General Theory of Relativity and Quantum Mechanics\footnote{In the present essay ``Quantum Mechanics'' refers to the standard (or textbook) formulation of quantum theory.} (QM), the pillars of contemporary physics, are our most accurate answers to the questions concerning the inherent nature of spacetime and matter respectively, changing the conception of the world provided by classical gravitational theories and Classical Mechanics (CM). Interestingly, Einstein's theory not only established that gravity is related to spacetime's curvature, but also that Euclidean geometry is not the correct mathematical theory in order to represent the physical world, being Riemannian geometry the appropriate framework to be adopted in General Relativity. Similarly, QM describes atoms and molecules and their dynamical behavior in a way that drastically changed our conception of reality provided by classical physics, speaking about intrinsically indeterminate systems, non-local interactions and superpositions; in addition, it even suggests a revision of classical logic. More precisely, the physical content of QM entails that the logical structure of the theory is not conformed to the laws of classical propositional calculus, as systematically shown for the first time in \cite{vonNeumann:1936}\footnote{The core idea of this paper, i.e.\ that the propositions of quantum theory do not conform to classical propositional logic, is also contained in embryonic form in von Neumann's monograph \emph{Mathematische Grundlagen der Quantenmechanik} appeared earlier in 1932. Interestingly, the idea that QM may implement logics different from the classical propositional calculus circulated before Birkhoff and von Neumann paper. For instance, in 1931 the Polish logician Zygmunt Zawirski proposed to apply a three-valued logic to quantum theory, starting from considerations about the wave-particle duality and Heisenberg's uncertainty principle. Furthermore, in 1933 Fritz Zwicky suggested that QM rejects the law of excluded-middle. For historical details see \cite{Jammer:1974}, Chapter 8. However, \cite{vonNeumann:1936} is generally considered the founding text of quantum logic, and here I will follow this tradition.}---such a result had a notable echo, opening new research programs concerning not only the logic of QM (cf.\ \cite{Reichenbach:1944}, \cite{Mackey:1957}, \cite{Finkelstein:1963}, \cite{Kochen:1965}, \cite{Jauch:1969}, see \cite{Jammer:1974} Chapter 8 for a historical perspective), but also regarding the nature of logic itself and its empirical status (cf.\ notably \cite{Quine:1951}, \cite{Putnam:1968}, \cite{Dummett:1976}, \cite{Hallett:1982}, \cite{Weingartner:2004},  \cite{Bacciagaluppi:2009}).\footnote{For space reasons I will not enter into the latter debate, which is orthogonal to the aim of the present essay.} 

In this paper we will be concerned with a precise question: does quantum physics necessarily implies a revision of classical logic? Alternatively stated, we will ask whether classical logic must be abandoned in the quantum realm or if there are quantum theories which allow to retain a classical propositional calculus.\footnote{In the present essay I will be concerned only with propositional calculus in the context of non-relativistic QM; thus, first order logic and higher order logics will not be discussed, as well as the logic of relativistic quantum theories.}
\vspace{2mm}

In order to begin our investigation, it is opportune to explain what is the relation between logic and mechanics, which is algebraic in essence. It is a well known fact that classical propositional logic is equivalent to a Boolean algebra where propositions (atomic and complex) have truth values ``true'' and ``false'' (or 1 and 0), and the main operations among them are conjunction ``$\wedge$'', disjunction ``$\vee$'' and negation ``$\neg$'', the only unary operation. From these logical connectives one can introduce secondary operations, as for instance the material implication ``$\longrightarrow$'', the exclusive or ``$\oplus$'' and logical equivalence ``$\longleftrightarrow$''.\footnote{For convenience, in this essay I will adopt the triple $(\wedge, \vee, \neg)$ as a basis of connectives. However, the choice of the basis is not unique.} Furthermore, it is important for our discussion to underline that the laws of propositional logic include commutativity, associativity, identity and distributivity with respect to $\wedge$ and $\vee$.

Analogously, the algebraic structure underlying classical mechanics is Boolean as well, being commutative, distributive and associative; in this context the logical operations of conjunction, disjunction and negation are replaced respectively by multiplication, addition and complementation among the variables of the theory. This fact has an interesting physical significance, since the variables associated to magnitudes of physical systems can be added and multiplied together---i.e.\ one can sum and multiply measurement results. Both propositional logic and classical mechanics, thus, generate a complemented Boolean lattice; from this fact it follows that the observable algebra of CM ``is isomorphic to a Boolean algebra of propositions with $(\wedge, \vee, \neg)$'' (\cite{David:2015}, p. 79).\footnote{Referring to this, Beltrametti more explicitly claims that ``Boolean algebras are algebraic models of classical logic (more specifically of classical propositional calculus) with the algebraic operations of meet ($\wedge$) and join ($\vee$) corresponding to the logical connectives ``and'', ``or'', and the unary relation of orthocomplementation corresponding to the logical negation. The rules and tautologies of classical logic have their counterpart in equations and identities of Boolean algebras'' (\cite{Beltrametti:2004}, p.\ 341).} Therefore, it is possible to formally characterize the state of a physical system via logical propositions which can assume the truth values ``true'' or ``false''---depending whether such statements describe true or false state of affairs concerning the system under consideration (cf.\ \cite{Jaeger:2009}, p. 61)---and to perform logical operations among propositions via the connectives $(\wedge, \vee, \neg)$. It is worth noting that in the context of CM it is always determined whether a system instantiates a given property or not; alternatively stated, every logical proposition about physical systems is either true or false, and hence in CM the principle of semantic bivalence holds (for more details cf.\ \cite{Bub:2007}, p.\ 642, \cite{Giuntini:2002}, p.\ 130).
 
Taking instead into account the mathematical structure and the physical content of standard QM, the classical propositional calculus is no longer appropriate to represent the logic of quantum propositions, since quantum theory does not share the same algebraic features of classical mechanics. In particular, the algebra of quantum observables is non-commutative and from a logical perspective generates a non-distributive orthocomplemented lattice, which is non-Boolean. Looking at the vast literature on quantum logic, it is possible to note that currently there exists a multitude of different quantum logical systems, but in this essay we will stick to the standard approach contained in \cite{vonNeumann:1936}, which will be introduced in the next section. 

Recalling what has been said above, the formal structure and the physical content of QM triggered vivid discussions about the necessity of a new kind of logic in order to understand what the theory says about the world, and hence, to comprehend how the microphysical realm behaves. Nonetheless, it should be noted in the first place that standard quantum theory is affected by severe conceptual and technical issues, e.g.\ the measurement problem, the lack of a physically satisfactory explanation of the wave function collapse, the presence of ill-defined terms within the axioms of the theory, \emph{etc}. (cf.\ \cite{Bell:2004aa}, \cite{Maudlin:1995aa}, \cite{Bricmont:2016aa}). Since Quantum Logic (QL) rests upon the very same formalism of QM, many authors argued that it does not have the necessary tools in order to provide solutions to such foundational issues, as for instance \cite{Bacciagaluppi:2009} and \cite{Giuntini:2002}.\footnote{In what follows I will not discuss the useful practical applications of QL in the field of quantum information and computation.} 
Referring to this, several strategies have been proposed to solve the conceptual and technical conundra of standard QM, e.g.\ to modify the formalism of the theory, a path followed by the spontaneous collapse theories (cf.\ \cite{Bassi:2003}), to provide new interpretations of the quantum formalism, as done by Everett's relative state formulation (\cite{Everett:1957aa}), the many world interpretation (\cite{Wallace:2012aa}) or more recently by Rovelli's relational mechanics (\cite{Rovelli:1996}), or to add variables to the quantum formalism, as done in Bohmian Mechanics (BM) (\cite{Durr:2013aa}). Here we will be concerned with the latter theory. 

More specifically, the principal aim of the present work is to show that it is not necessarily true that QL is indispensable in \emph{every} quantum theory, or better, that quantum physics \emph{necessarily} entails a revision of classical logic. It is \emph{not} my intention to claim that QL is not needed (or it is useless) to account for the physical content of standard non-relativistic QM, but rather, I will challenge that a new quantum logic is indispensable in order to model the propositions in the context of every conceivable quantum theory. In this paper it will be showed, indeed, that taking into account BM, a theory implementing an ontology of particles in motion in three-dimensional physical space, it is possible to restore a classical interpretation of the logical connectives. The classical logical structure of such a theoretical framework, it will be argued, is another remarkable consequence of the clear ontology of the theory, a fact which is not yet sufficiently recognized in the literature concerning BM. The present work, then, aims to fill this gap. 

The essay is structured as follows: in Section \ref{QL} I will introduce the standard approach to quantum logic introduced by Birkhoff and von Neumann, whereas in Section \ref{BM} Bohmian mechanics will be presented in some detail, focusing in particular on its metaphysics and its theory of measurement. In this manner, I will be able to explain in Section \ref{CL} the reasons for which in BM logical connectives retain their classical interpretation, showing that there exists a quantum theory with an underlying classical logical structure. This fact, in turn, implies that the universal claim for which quantum physics entails a revision of classical logic is not always true. Section \ref{conc} concludes the paper.

\section{Quantum Logic in a Nutshell}
\label{QL}

The principal aim of Birkhoff and von Neumann's founding essay of quantum logic was to ``discover what logical structures one may hope to find in physical theories which, like quantum mechanics, do not conform to classical logic'' (\cite{vonNeumann:1936}, p.\ 823). The expression quantum logic in this paper must be clarified, since it refers to a quantum propositional calculus in the form of a ``calculus of linear subspaces with respect to \emph{set products}, \emph{linear sums}, and \emph{orthogonal complements}'' which ``resembles the usual calculus of propositions with respect to \emph{and}, \emph{or}, and \emph{not}'' (\emph{ibid}.), where the logical propositions are associated to measurements, tests on quantum systems.\footnote{Cf.\ \cite{Giuntini:2002} and \cite{DallaChiara:2004} for a systematic introduction to various forms of quantum logics, and to \cite{Engesser:2009} for historical and philosophical discussions on the topic.}
Let us now explain the reasons for which quantum theory does not conform to classical logic, recalling some of its well-known mathematical features.\footnote{For details on the mathematical structure of QM and its physical content the reader may refer e.g.\ to \cite{Sakurai1994}, and \cite{Griffiths:2014}. In this essay, it is assumed that the reader has some familiarity with QM and QL.} 

In the first place, QM relies on a non-commutative algebraic structure, contrary to the case of classical mechanics. Indeed, quantum observables\footnote{In QM properties of physical systems are represented by Hermitian operators $A$ called \emph{observables}. The possible states in which the system can be found after the performance of measurement are the eigenstates of $A$, and the corresponding values are the eigenvalues of $A$.} generally do not commute, i.e.\ for any pair of operators $A,B$ we have $[A,B]=AB-BA\neq0$. From a physical perspective this fact entails that if one performs a measurement of $A$ followed by a measurement of $B$ on a quantum system, in general one will obtain different results inverting the order of these observations. Moreover, in virtue of the non-commutative structure of quantum theory, to measure an operator $C=A+B$ is not typically equivalent to measure $A$ and $B$ independently and then add the respective results, since $A,B$ will be in general incompatible (cf.\ \cite{David:2015}, p.\ 77). This algebraic fact of QM constitutes the formal basis to prove the Heisenberg uncertainty relation, a theorem of quantum theory which reflects the operational inability to simultaneously measure the values of incompatible operators with arbitrary precision. Such a result, in turn, is usually interpreted ontologically in the sense that quantum system do not instantiate definite properties in non-measurement situations (cf.\ \cite{Sakurai1994}).

In the second place, another remarkable difference with respect to classical mechanics is that quantum systems can be in superposition states as a consequence of the linearity of the Schr\"odinger Equation (SE)---the fundamental dynamical law of QM---which for a single particle reads: 
\begin{align}
\label{SE}
i\hbar\frac{\partial\psi}{\partial t}=\Big(-\frac{\hbar^2}{2m}\nabla_k^2+V\Big)\psi=H\psi,
\end{align}
\noindent where $H$ represents the Hamiltonian operator, defined as the sum of kinetic and potential energy of the system at hand. More specifically, this algebraic property of \eqref{SE} entails that if two wave functions\footnote{In QM the wave function of a system provides the maximal information available about it.} $\psi_1, \psi_2$ are both possible solutions of the same Sch\"odinger equation, then their linear combination (\emph{superposition}) 
\begin{align}
\label{SSE}
\psi_s=\alpha\psi_{1}+\beta\psi_{2}
\end{align} 
is still a solution of the same SE---$|\alpha|^2,|\beta|^2$ (with $\alpha,\beta\in\mathbb{C}$) represent the probabilities to find the system in $\psi_1, \psi_2$ respectively. Notably, the new superposed state $\psi_s$ is also a consistent representation of the system.  Since in experimental situations such superpositions are never revealed, QM prescribes that the interaction between the measured system and the measurement apparatus causes the suppression of the Schr\"odinger evolution, collapsing stochastically the wave function in one of the possible eigenstates of the measured observable. The probability to find the system in one of such states is given by the Born's rule. It is in the measurement process, then, that the inherently stochastic character of quantum theory emerges vigorously, denying any form of causality or determinism. 

Given the axioms of QM, Birkhoff and von Neumann showed how the physical content of this theory does not conform to a classical logical structure. Their analysis begins by defining quantum logical propositions as \emph{experimental propositions}, i.e.\ statements affirming that a certain observable $A$ (or a set of observables) measured on a quantum system has a given value $a_i$ (or a sequence of values).\footnote{More precisely, the authors stress that both in classical and quantum mechanics observations of physical systems are given by readings of experimental outcomes ($x_1, \dots, x_n$) of \emph{compatible} measurements ($\mu_1, \dots, \mu_n$). The values ($x_1, \dots, x_n$) are elements of what the authors called the ($x_1, \dots, x_n$)-space, i.e.\ the ``observation-space'' of the system in question, whose elements are all the possible combinations of results of the measurements ($\mu_1, \dots, \mu_n$); thus, the actual values ($x_1, \dots, x_n$) form a subset of such a space. Birkhoff and von Neumann, then, defined the experimental propositions concerning a physical system as the subsets of the observation space associated with it.}

Secondly, the authors underlined that in classical and quantum mechanics the states of physical systems are mathematically represented by points in their state spaces---phase space for the classical case, and Hilbert space $\mathcal{H}$ for the quantum case---which provide the maximal information concerning the system. A point in phase space corresponds to the specification of the position and momentum variables of a certain classical system, whereas in QM the points of $\mathcal{H}$ are wave functions. The authors, then, find a connection between subsets of the observation-space of a system and subsets of the Hilbert space, specifying that quantum experimental proposition are mathematically represented by a closed linear subspace of $\mathcal{H}$; this step is crucial in order to obtain the quantum propositional calculus. More precisely, a proposition $p$ is associated with a subset $P\subset\mathcal{H}$ whose elements are all the pure states for which $p$ is true, hence, quantum logical propositions ``are in bijective correspondence to closed subspaces of the Hilbert space of the system'' (\cite{Bacciagaluppi:2009}, p.\ 55). 
Alternatively, we can say that quantum mechanical operators correspond to propositions with ``yes/no'' (``true/false'') outcome in a logical system as underlined by \cite{David:2015}, p.\ 78: 
\begin{quote}
An orthogonal projector $\textbf{P}$ onto a linear subspace $P\subset\mathcal{H}$ is indeed the operator associated to an observable that can take only the values 1 (and always 1 if the state $\psi\in P$ is in the subspace $P$) or 0 (and always 0 if the state $\psi\in P^{\perp}$ belongs to the orthogonal subspace to $P$). Thus we can consider that measuring the observable $\textbf{P}$ is equivalent to perform a test on the system, or to check the validity of a logical proposition $p$ on the system.
\end{quote}

In order to properly define a propositional calculus for QM, one must define the logical operators for conjunction, disjunction and negation and the notion of logical implication. Following Birkhoff and von Neumann, the procedure is rather simple: 
\begin{enumerate}
\item The negation of a proposition $p$ is defined by the authors as follows: ``since all operators of quantum mechanics are Hermitian, the mathematical representative of the negative of any experimental proposition is the orthogonal complement of the mathematical representative of the proposition itself'' (\cite{vonNeumann:1936}, pp.\ 826-827). The orthogonal complement $P^{\perp}$ of the subspace $P$ is the set whose elements are all the vectors orthogonal to the elements of $P$. From a physical perspective such an orthogonal complement satisfies the following property: given a subset $P\subset\mathcal{H}$ and a pure state $\psi$, $\psi(P)=1$ iff $\psi(P^{\perp})=0$ and  $\psi(P)=0$ iff $\psi(P^{\perp})=1$. As Dalla Chiara and Giuntini underline, ``$\psi$ assigns to an event probability 1 (0, respectively) iff $\psi$ assigns to the orthocomplement of $P$ [notation adapted] probability 0 (1, respectively). As a consequence, one is dealing with an operation that \emph{inverts} the two extreme probability-values, which naturally correspond to the truth-values \emph{truth} and \emph{falsity} (similarly to the classical truth-table of negation)'' (\cite{Giuntini:2002}, p.\ 132).
\item Concerning the conjunction, Birkhoff and von Neumann notice that one can retain the very same set-theoretical interpretation as the classical conjunction, since the intersection of two closed subspaces $P, Q\subset\mathcal{H}$ is still a closed subspace. Thus, one maintains the usual meaning for the conjunction: the pure state $\psi$ verifies $P\cap Q$ iff $\psi$ verifies both $P$ and $Q$. Thus, quantum logic does not introduce a new logical operator for the conjunction.
\item Contrary to the previous case, the logical operator for disjunction cannot be represented by the set-theoretic union, since the set-theoretical union of the two subspaces $P, Q$ will not be in general a subspace, thus, it is not an experimental proposition. Therefore, in QL one introduces the quantum logical disjunction as the the closed span of the subspaces $P, Q$, which is an experimental proposition. Such a statement corresponds to the ``smallest closed subspace containing both $P$ and $Q$'' (\cite{Bacciagaluppi:2009}, p.\ 55). 
\item The logical implication is defined by set theoretical inclusion: given two experimental propositions $p$ and $q$, 
$p$ implies $q$, means that whenever one predicts $p$ with certainty, one can predict also $q$ with certainty, and this is equivalent to state that $p$ is a subset of $q$. This fact is particularly important since the authors showed that ``it is \emph{algebraically} reasonable to try to correlate physical qualities with subsets of phase-space'' and thus, ``physical qualities attributable to any physical system form a partially ordered system'' (\cite{vonNeumann:1936}, p.\ 828, notation adapted).
\end{enumerate}

In this manner we have defined a propositional calculus for quantum experimental propositions, which generates an orthocomplemented lattice. The remarkable fact about this quantum logic---and the crucial difference with respect to the logic associated with experimental propositions of classical mechanics---is the failure of the distributive law. It is a well-known fact that such a law holds in the propositional calculus underlying classical mechanics in virtue of the commutative algebra of classical observables. However, conjunction and disjunction are not distributive in the context of QL, meaning that given three propositions $p, q$ and $r$ the logical law
\begin{align}
\label{DL}
p\wedge(q\vee r)\longleftrightarrow(p\wedge q)\vee(p\wedge r)
\end{align}

\noindent does not hold.\footnote{In the context of quantum logic, distributivity must be replaced by a weaker law: $(p\wedge q)\vee(p\wedge r)\longrightarrow(p\wedge(q\vee r)).$}
Let us explain this fact with a simple example taken from \cite{Giuntini:2002}. These authors claim that distributivity fails in quantum logic because of (i) the peculiar behavior of the quantum disjunction---which may be true although neither of the member is true in virtue of the linearity of \eqref{SE}---and (ii) the non-commutativity of quantum observables. More concretely, let us suppose to consider a $1/2-$spin particle, let's say an electron that has spin up along the $x$ axis. From QM we know that there are only two possible spin states in which a particle can be found in each axis after a spin measurement, namely either in the up or in the down state. Furthermore, since spin operators along different axis do not commute, being incompatible observables---$[S_x, S_y]=[S_x, S_z]=[S_y, S_z]\neq0$---they cannot be simultaneously measured. Then, as a consequence of the Heisenberg uncertainty principle, ``both propositions ``$spin_y$ is up'' and ``$spin_y$ is down'' shall be strongly undetermined. However, the disjunction ``either $spin_y$ is up or $spin_y$ is down'' must be true'' (\cite{Giuntini:2002}, pp.\ 133-134), and the same holds for the case of spin in the $z$ direction. Therefore, if $p$ in \eqref{DL} represents the proposition ``the electron has $x$-spin up'', $q$ the proposition ``the electron has $y$-spin up'' and $r$  the proposition `the electron has $y$-spin down'', it is possible to see that the first part of \eqref{DL} express a true statement---i.e.\ $p\wedge(q\vee r)=1$---whereas $(p\wedge q)$ and $(p\wedge r)$ are both false in virtue of the incompatibility between the spin operators among different axes, implying that the second part of the distributive law asserts a false statement, i.e.\ $(p\wedge q)\vee(p\wedge r)=0$. Consequently, this law fails in the context of quantum theory, since we have $p\wedge(q\vee r)=1\longleftrightarrow(p\wedge q)\vee(p\wedge r)=0$. 
\vspace{2mm}

As we have seen, Birkhoff and von Neumann claimed that a quantum logic should be introduced since the physical content of QM implies a violation of classical logical laws and a redefinition of some logical operations among propositions. After the publication of this paper, as already stated in the previous section, research in quantum logic gathered notable attention and it is now a well-established sub-discipline in the foundational research in quantum physics.\footnote{For historical clarity it must be said that Birkhoff and von Neumann's paper did not obtain much attention immediately after its publication, but it became a central work in quantum foundations from the fifties.} However, it must be underlined that QL, resting on the mathematical structure of QM, cannot solve the cogent problems which affect quantum theory itself, as for instance the quantum measurement problem, or the incompatibility among the dynamical laws of the theory. This conclusion is now widespread among experts as exemplified e.g.\ by \cite{Bacciagaluppi:2009} and \cite{Giuntini:2002}. Referring to this, the latter authors stated explicitly that ``quantum logics are not to be regarded as a kind of ``clue'', capable of solving the main physical and epistemological difficulties of QT [quantum theory]. This was perhaps an illusion of some pioneering workers in quantum logic'' (\cite{Giuntini:2002}, p.\ 225). 

In the remainder of this essay, we shall concentrate on Bohmian mechanics, a quantum theory where the usual problems of non-relativistic QM are solved via the introduction of a clear particle ontology. It will be shown that in virtue of such a metaphysical picture, this theory presents a classical logical structure. Thus, we will see that quantum physics does not necessarily entails a revision of classical propositional calculus.

\section{Bohmian Mechanics, Observables and Quantum Measurements}
\label{BM}

Bohmian mechanics is a quantum theory implementing a Primitive Ontology (PO) of point-particles moving along deterministic trajectories in physical space. The primitive ontology of a theory $T$ defines the fundamental entities of $T$, those objects which cannot be further analyzed and/or defined in terms of more elementary notions contained in the theory's vocabulary. These entities are the variables appearing in $T$'s equations provided with a \emph{direct} physical meaning, i.e.\ referring to \emph{real} objects precisely localized and moving in physical space, and explain---together with the dynamical laws of the theory under consideration---the macroscopic world surrounding us. Thus, as we will see more clearly later on, physical phenomena are explained in BM exclusively in terms of particles in motion in physical space (and ontologically reduced to them).\footnote{The interested reader may refer to \cite{Allori:2013ab} and \cite{Esfeld:2014ac} for a more detailed explanation of the notion of primitive ontology.} For the sake of precision and simplicity, in the remainder of the present work I will focus exclusively on the version of BM contained in \cite{Durr:2009fk} and \cite{Durr:2013aa}, where the Bohmian particles are the only entities defined in space and time, and the quantum mechanical wave function is considered an abstract nomological entity, not living in 3-dimensional space.\footnote{Other versions of the theory maintain different positions concerning the metaphysical status of the $\psi$-function; see notably \cite{Bohm:1952aa}, \cite{Holland:1993} and \cite{Romano:2018}.} In what follows, I will show that in virtue of its clear particle ontology such a theoretical framework implements a classical logical structure. Alternatively stated, the classical logical structure of BM is a consequence of its primitive ontology. 
\vspace{2mm}

Bohmian mechanics provides a complete description of quantum systems via the introduction of additional variables to the wave-function, i.e.\ the specification of particles' positions. In this theory physical systems are represented by a couple $(Q(t), \psi(t))$, where $Q(t)=(Q_1(t),\dots, Q_N(t))$ describes the actual positions in space of a $N$-particle configuration at an arbitrary time $t$, and the latter is the configuration's wave function, which does not exhaust the information concerning physical systems as in QM. Bohmian particles are always precisely localized, hence, the corpuscles follow trajectories in physical space. Referring to this, the dynamics of BM is fully described by two equations of motion, one for the $\psi$-function, one for the Bohmian particles. The former evolves in $3N$-configuration space according to the unitary evolution provided by \eqref{SE}, whereas the latter are governed by the so-called \emph{guidance} equation:

\begin{align}
\label{guide}
\frac{dQ_k}{dt}=\frac{\hbar}{m_k}\mathrm{Im}\frac{\psi^*\nabla_k\psi}{\psi^*\psi}(Q_1,\dots, Q_N)=v_k^{\psi}(Q_1,\dots, Q_N).
\end{align}
 
\noindent From \eqref{guide} it is clear that the particles' velocity is a function of $\psi$: the guidance equation generates a vector velocity field (see r.h.s.) which depends on the wave function, whose role is to guide the motion of the particles in physical space on the one hand, and to determine the statistical distribution of the particles' positions on the other. Remarkably, since the velocity of the particles is defined as the rate of change in time of their position, it is a secondary property which can be ascribed to the Bohmian corpuscles.\footnote{It is worth stressing that although BM implements a corpuscular ontology, it is a non-classical theory since it is explicitly non-local, as one may see from the r.h.s. of \eqref{guide} (cf.\ \cite{Durr:2009fk}, Chapter 8, \cite{Bricmont:2016aa}, Chapter 5).}

Finally, the Born's distribution is preserved in BM, making its predictions empirically indistinguishable with respect to those of QM. More precisely, to understand this result one should take into consideration a crucial consequence of the Bohmian dynamics, i.e.\ equivariance: if  at an arbitrary initial time the initial particle distribution is given by the density $\rho_0(Q)=|\psi(Q, 0)|^2$, the density at later times will be given by $\rho_t(Q)=|\psi(Q, t)|^2$.\footnote{For the mathematical justification of this statement the reader should refer to \cite{Durr:2013aa}, Chapter 2, Secs. 4-7. To this regard, it is also worth noting that BM restores also \emph{a classical interpretation of quantum probabilities}. These are manifestation of our ignorance about the exact positions of the Bohmian particles and our operational inability to manipulate them. Thus, the maximum knowledge of particles' configurations at our disposal in BM is provided by $|\psi|^2$.} 

\subsection{The Bohmian Treatment of the Measurement Process}

In order to present the Bohmian theory of measurement in a straightforward manner, let us consider the very simple, idealized case of a wave function $\psi$ in a superposition of two states $\psi_1, \psi_2$ corresponding to the possible eigenstates of a two-valued operator $O$, with eigenvalues ``left'' ($L$) and ``right'' ($R$):
\begin{align}
\psi=c_1\psi_{1}+c_2\psi_{2}, 
\end{align}
\noindent where $c_1, c_2\in\mathbb{C}$ and $|c_1|^2+|c_2|^2=1$. Before the measurement's performance, we assume that the macroscopic experimental device used to measure $O$ is in the ready state $\Phi_0$, i.e.\ in a state pointing to a neutral direction, whereas the other admissible pointer's positions will be $\Phi_1$ for $L$ (pointing to the left) and $\Phi_2$ for $R$ (pointing to the right). From ordinary quantum theory we know that  system and apparatus are initially independent, i.e.\ described by a product wave function; however, given the deterministic evolution provided by the Schr\"odinger equation we obtain a macroscopic superposition:
\begin{align}
\label{macrosuperpos}
\sum_{i=1,2}c_i\psi_i\Phi_0\longrightarrow\sum_{i=1,2}c_i\psi_i\Phi_i.
\end{align}


To tame this awkward feature of standard QM---which is the quantum measurement problem mentioned in the previous sections---Bohmian mechanics provides a description of the measurement process by specifying the trajectories of the individual particles composing both the system and the classical apparatus. Indeed, according to this theoretical framework a measurement is just an interaction between the system and the experimental device, and both are composed of localized particles which evolve according to the laws \eqref{SE}-\eqref{guide}. 

Notably, in this theory a particular measurement situation is formally represented by a couple $(X,Y)\in\mathbb{R}^{3M}\times\mathbb{R}^{3(N-M)}$ where the former variable refers to the actual initial $M$-particle configuration of the subsystem under consideration, and the latter to the actual configuration of particle composing the environment, i.e.\ its complement formed by all the particles not in the subsystem, including the particle configuration of the experimental device which will actually register the measurement result. 
BM provides an algorithm to define the \emph{effective wave function} of the subsystem's degrees of freedom, which corresponds to the usual quantum mechanical wave function.\footnote{For technical details see \cite{Durr:2013aa}, Chapter 2, and \cite{Durr:2009fk}, Chapter 9.} In a nutshell, the system's effective wave function $\psi_t(x)$ is obtained from the universal wave function $\Psi=\Psi(x,y)$---i.e.\ the wave function of the particle configuration composing the entire universe---where $x,y$ represent generic variables for the subsystem and its complement respectively. The next step to define $\psi_t(x)$ is to insert the \emph{actual} configuration of the environment at hand $Y_t$ into the wave function $\Psi(x,y)$. In this manner we obtain the so-called \emph{conditional wave function} $\psi_t(x)=\Psi(x_t, Y_t)$ of our subsystem's degrees of freedom; such a function obeys the Schr\"odinger equation only under the particular condition in which the system is decoupled from its environment. This entails that the larger wave function must have the following effective product form:
\begin{align}
\Psi(x, y)=\psi(x)\Phi(y)+\Psi^{\bot}(x, y) 
\end{align}

\noindent with $\Phi(y)$ and $\Psi^{\bot}(x, y)$ having macroscopically disjoint supports\footnote{The support of a wave function is the domain in which it is non-zero. ``The notions of support, separation of supports, and disjointness of supports have to be taken with a grain of salt. The support of a Schr\"odinger wave function is typically unbounded and consists of (nearly) the whole of configuration space. ``Zero'' has thus to be replaced by ``appropriately small'' (in the sense that the square norm over the region in question is negligible). Then, the precise requirement of macroscopic disjointness is that the overlap of the wave functions is extremely small in the square norm over any macroscopic region'', \cite{Durr:2009fk}, p. 174, footnote 1. In this regard, it is interesting to note that the pointer's wave function ``is a macroscopic wave function, which can be imagined as a ``random'' superposition of macroscopically many ($\approx 10^{26}$) one-particle wave functions, with support [footnote deleted] tightly concentrated around a region in configuration space (of $\approx 10^{26}$ particles) that makes up a pointer in physical space pointing in some direction, i.e., defining some pointer position. So different pointer positions belong to macroscopically disjoint wave functions, that is, wave functions whose supports are macroscopically separated in configuration space'' (\emph{ibid.}).} in the $y$-variables and $Y_t\in\mathrm{supp}\Phi(y)$. Thus, for all practical purposes one may not consider the empty branch of the wave function $\Psi^{\bot}$. When this condition is met, the subsystem's conditional wave function is called \emph{effective wave function}.\footnote{Referring to this, Passon notes that in measurement situations ``the wavefunction of the measurement apparatus will in general be in a superposition state. The configuration however indicates the result of the measurement which is actually realized. That part of the wavefunction which ``guides'' the particle(s) can be reasonably termed the \emph{effective wavefunction}. All the remaining parts can be ignored, since they are irrelevant for the particle dynamics. As a result of decoherence effects [...], the probability that they will produce interference effects with the effective wavefunction is vanishingly small'' (\cite{Passon:2018}, p. 191).}

Having explained how BM defines subsystems' wave functions, let us return to our example taking into account Figure 1, which represents the three possible orientations of the macroscopic pointer in physical space, and the relative support of the pointer's wave function in configuration space described a few lines above.

\begin{center}
\includegraphics[scale=0.6]{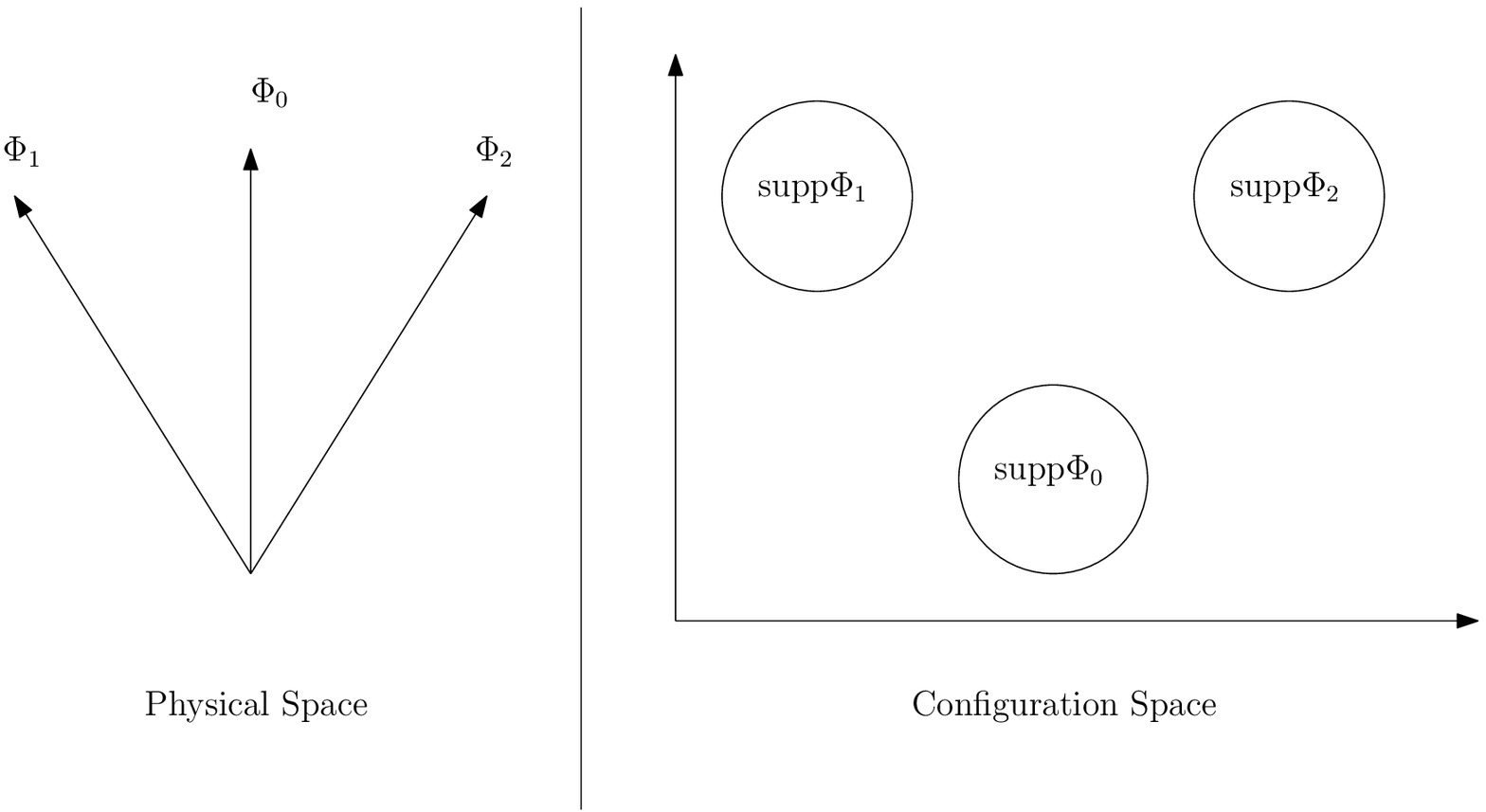}
\begin{quote}
\footnotesize{Fig. 1: Schematic representation of the three different possible directions of the pointer in physical space (left) and the relative supports of the pointer's wave function in configuration space (right).}
\end{quote}
\end{center}

\noindent As already stated, before the performance of the measurement at time $t=0$, the macroscopic pointer points in the neutral direction in the ``ready'' state and, consequently, the wave function of the actual configuration of particle composing the experimental device has support $Y_0\in\mathrm{supp}\Phi_0$. At this stage the system has not yet interacted with the experimental device, thus, they are initially independent and described by a product wave function.
The two other pointer position wave functions, corresponding to the possible outcomes $L$ and $R$ respectively, are $\Phi_1(y)$ where $Y_t\in\mathrm{supp}\Phi_1$, and $\Phi_2(y)$, $Y_t\in\mathrm{supp}\Phi_2$, as shown in Figure 2.

\begin{center}
\includegraphics[scale=0.6]{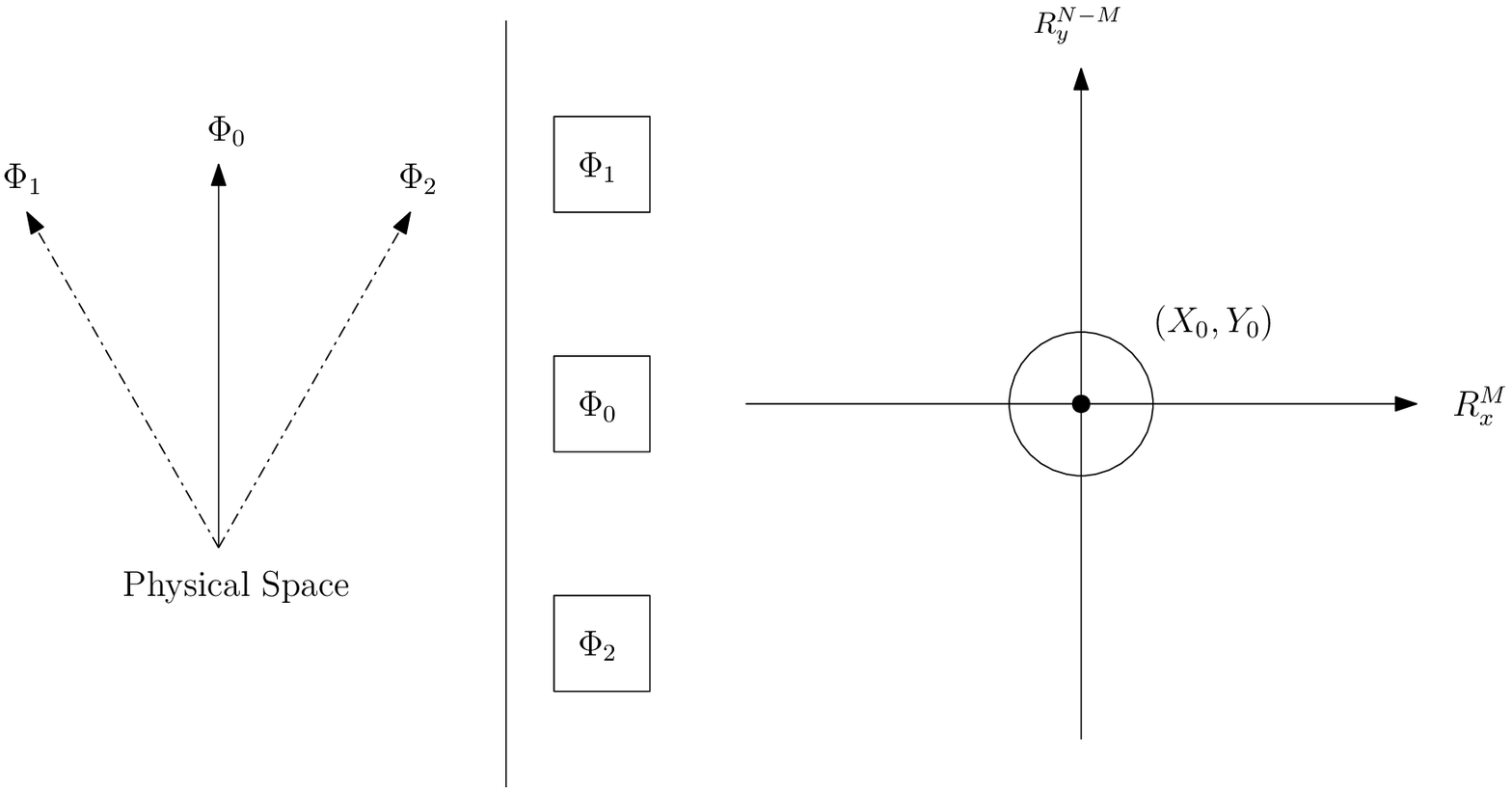}
\begin{quote}
\footnotesize{Fig. 2: Schematic representation of the pointer pointing in the neutral direction in physical space before the measurement, solid line. Dashed lines represent physically possible but not actualized measurement outcomes (left). The relative support of the pointer's wave function and particle configuration describing the experimental situation before the measurement's performance (right).}
\end{quote}
\end{center}

\noindent Running the experiment, the interaction between the system and the experimental device will be governed exclusively by the dynamical evolution provided by \eqref{SE} and \eqref{guide}. Since particles' positions of the measured system and the experimental apparatus are always well-defined, the measurement outcome will be also well-defined as a consequence of the theory's formalism, i.e.\ the final outcome is a function of (i) the initial conditions of the particles composing the system and the experimental device, and (ii) the dynamical laws governing their behavior. Hence, the configuration of particles $(X_0,Y_0)$ will deterministically evolve at a successive time $t>t_0$ into another configuration $(X_t,Y_t)$, corresponding to one of the possible eigenstates of the measured operator $O$, coupled with a definite state of the macroscopic experimental device, which will point in a definite direction expressed explicitly by $Y_t$.\footnote{For technical details about the Bohmian theory of measurement the reader may consult \cite{Durr:2009fk}, Chapter 9, \cite{Bohm:1952aa}, Part II, Section 2 and \cite{Maudlin:1995ab}.} More precisely, after the measurement, $Y_t$ will have a well-defined macroscopic support in the $y$-variable, i.e.\ either we obtain $Y_t\in\mathrm{supp}\Phi_1(y)$ or $Y_t\in\mathrm{supp}\Phi_2(y)$, and the empty packet of the wave function can be practically neglected. Suppose to find the outcome $R$, then, the macroscopic pointer will point in the $\Phi_2$ direction, implying that $Y_t\in\mathrm{supp}\Phi_2(y)$, as illustrated in Figure 3.

\begin{center}
\includegraphics[scale=0.7]{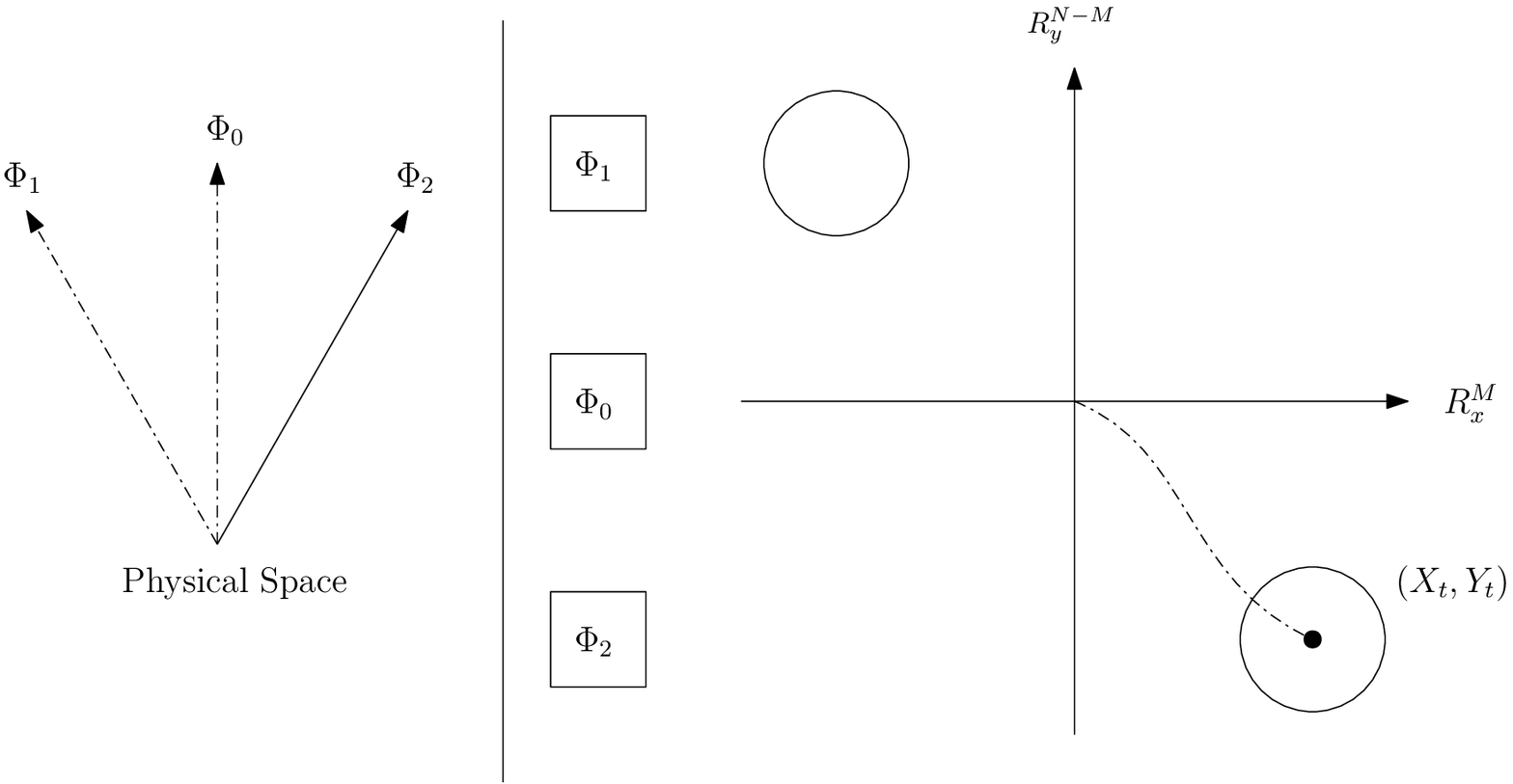}
\begin{quote}
\footnotesize{Fig. 3: Schematic representation of the pointer pointing in the right direction in physical space, meaning that the outcome $R$ has been obtained, solid line (left). The relative support of the pointer's wave function and particle configuration describing the experimental result are shown on the right. For practical purposes the empty branch of the wave function can be neglected.}
\end{quote}
\end{center}

Hence, according to BM, measurement results are simply defined as functions of the primitive ontology of the theory and its dynamics. As \cite{Goldstein:2012} said, ``a statement like ``the experiment $E$ has the outcome $z$'' should mean that the PO of the apparatus indicates the value $z$. For example, if the apparatus displays the outcome by a pointer pointing to a particular position on a scale, what it means for the outcome to be $z$ is that the matter of the pointer is, according to the PO, in the configuration corresponding to $z$. Thus, the outcome $Z$ is a function of the PO, $Z=f(PO)$''. 

From our simple presentation of the Bohmian theory of measurement, four remarks must be emphasized:

\begin{enumerate}
   \item In this theory there are no superpositions of particles in physical space, therefore the physical situation described in \eqref{macrosuperpos} is avoided by construction. Consequently, macroscopic superpositions are forbidden;
   \item Measurement results are functions of the primitive ontology and its dynamical evolution provided by the Schr\"odinger equation and the guiding equation. Thus, the individual physical processes responsible for the macro-objectification of measurement outcomes are independent from external observers;
   \item Wave functions are not subjected to stochastic jumps: the wave function's collapse loses its fundamental role in the dynamics of the theory, being an \emph{effect} of the interaction between subsystem and environment. Hence, in BM observers never cause the result of a given measurement.\footnote{For significant literature about the role of observers in BM the reader may refer to \cite{Bohm:1952aa}, \cite{Durr:2013aa}, \cite{Durr:2009fk}, Chapter 9, \cite{Bricmont:2016aa}, Chapter 5, \cite{Passon:2018}.} 
\item Operators are associated with experiments, not with genuine properties of physical systems, since every measurement outcome depends on the initial positions of the Bohmian particles and the laws \eqref{SE}-\eqref{guide}. Therefore, in BM only particles position are really observed (\cite{Bell:1982aa}, p.\ 996).
\end{enumerate}

\subsection{Contextuality in Bohmian Mechanics}

In the light of what has been stated above, we can say that in the context of BM the usual quantum observables are reduced to position measurements; more precisely, observational outcomes are ontologically dependent on the initial conditions of the experimental set-up---i.e.\ the initial particle configuration of the measured system and the apparatus---and the dynamics of the theory (cf.\ \cite{Durr:2004c} in particular). It is not surprising, then, that BM is also a contextual theory: quantum observables are not considered metaphysically genuine properties of the particles, and their measurements do not reveal pre-existing values of such magnitudes. This theory, therefore, conforms to the no-go theorems showing the contextuality of quantum observables and the impossibility to complete QM with non-contextual hidden variables (cf.\ \cite{Lazarovici:2018}). Let us explain the latter point with an explicit example, taking into account a spin measurement on a single particle along the $z$-axis. 

Before the measurement of the $z-$spin, the system will be in general described by the following state:
\begin{align}
\psi(z)\Big(|\uparrow\rangle+|\downarrow\rangle\Big)
\end{align}

For the sake of simplicity I will assume that the wave function written above is symmetrical, i.e.\ $\psi(z)=\psi(-z)$, so that the nodal line showed in Figure 4 corresponds to the case in which $z=0$; such a nodal line cannot be crossed by Bohmian particles in virtue of \eqref{guide}. In the figure below, $H$ indicates the direction of the inhomogeneous magnetic field of the idealized Stern-Gerlach apparatus we are considering, the circles represents schematically---and in a very idealized way---the supports of the wave function. 

\begin{center}
\includegraphics[scale=0.7]{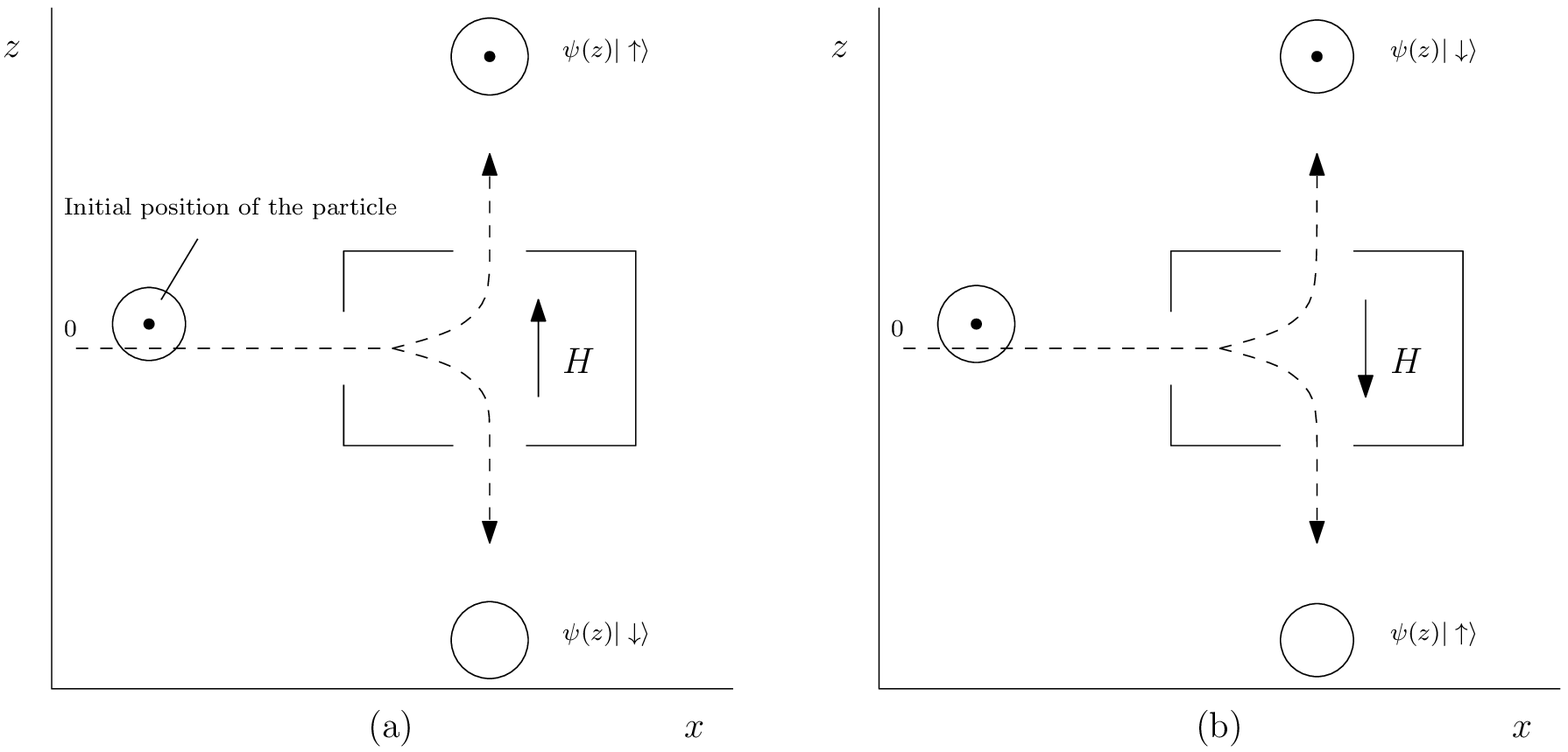}
\begin{quote}
\footnotesize{Fig. 4: Schematic representation of the contextual nature of the property of spin in Bohmian Mechanics. These picture are taken from \cite{Bricmont:2018}.}
\end{quote}
\end{center}

In this scenario if we perform a spin measurement along the $z$ axis we can obtain two results: one given by $|\uparrow\rangle$, which in figure 4a follows the direction of $H$, and one given by $|\downarrow\rangle$, which goes in the opposite direction. In our example, at the beginning of the measurement the Bohmian particle is located above the nodal line, and therefore it necessarily will go up, obtaining the state $\psi(z)|\uparrow\rangle$. Then, we will say that the particle ``has $z-$spin up''. Remarkably, if we reverse the direction of $H$ leaving unaltered the initial position of the particle above the nodal line, as in figure 4b, we will conclude that the particle will go up as well. However, this time the particle travels in the opposite direction with respect to the $H$ field, and we will say that the particle ``has $z-$spin \emph{down}''. According to BM, thus, the initial configuration of the particle and the measuring apparatus---in combination with the dynamics of the theory---will univocally determine the outcome. Hence, this measurement will not measure any pre-existing spin property of the particle under consideration. Notably, in BM this conclusion has been generalized to every quantum observable.

In sum, from our discussion it is possible to conclude that in the context of BM, every proposition about quantum measurements must be translated into a statement concerning the particles' positions and their dynamics. This fact is crucial to demonstrate that  the logical structure underlying Bohmian mechanics is classical.

\section{Classical Logic in Quantum Context}
\label{CL}

In the previous section we showed that in Bohmian mechanics measurement results are defined as function of the PO of the theory. This fact allows us to leave aside considerations about observables, since they do not refer to genuine properties of quantum systems---contrary to the cases of classical and quantum mechanics---being merely associated with experiments. Therefore, we need to concentrate uniquely on the particles' positions, the exclusive genuine (and the only measurable) property of Bohmian corpuscles. This entails, in turn, that logical (or experimental) propositions must be statements concerning the locations of particles and nothing else, given that such entities are the unique objects of the theory defined in physical space. This fact marks an important difference also with respect to classical mechanics, since in that context logical propositions assign a set of properties which are considered genuine attributes of physical systems, although they are function of positions and momenta. On the contrary, in BM suggests a minimal ontological commitment towards particles' position and their change in time, being particles' location the unique inherent properties of such entities (cf.\ \cite{Goldstein:2005a, Goldstein:2005b}, \cite{Esfeld:2015aa}).

In addition, one must also define the truth conditions for logical statements in BM, i.e.\ under what conditions a certain proposition is true or false. It is worth recalling that in this theory objects are completely determined, meaning that there is an objective fact of the matter---completely independent from external observers---establishing whether a certain particle instantiates a given property. 
Since in this theoretical framework every physical fact and phenomenon is reduced ontologically to the motion of the Bohmian particles in space, we will say that a logical proposition $p$ has truth value ``true'' if and only if there exists a physical state of affairs that makes $p$ true; a proposition is false otherwise. NB: given that in BM there are no superposition of particles, this theoretical framework maintains the principle of semantic bivalence, since the corpuscles are precisely and uniquely localized in space. To argue that BM has a classical logical structure, then, it is sufficient to show that in such a theory the logical connectives $(\wedge, \vee, \neg)$ maintain their classical interpretations. Let us then consider the following cases:
\begin{itemize}
\item \textbf{Classicality of} $\wedge$: Given two atomic propositions $p, q$, the complex proposition $p \wedge q$ is true in BM iff $p$ and $q$ are both true; this statement is false otherwise. Consequently, a proposition with $n$ conjuncts $p_1\wedge p_2\wedge\dots\wedge p_n$ is true if and only if every $p_i$ is true, and false otherwise (i.e.\ there must be a state of affairs that makes every conjunct true). Hence, the logical operator $\wedge$ retains its classical meaning. For instance, the sentence ``the particle $k$ has position $q_k$ and velocity $v_k$'' is true in BM iff it is the case that the particle $k$ is actually located in space at the position $q_k$ and if it has velocity $v_k$. If one of these two statements is false, the conjunction will be false as well. This case is not particularly interesting, since also in standard QL the conjunction is defined as in classical logic, so the meaning of this connective remain unaltered from the classical to the quantum transition.
\vspace{2mm}

\item \textbf{Classicality of} $\vee$: Given two atomic propositions $p, q$, the complex proposition $p\vee q$ is true in BM iff at least either $p$ or $q$ is true, i.e.\ there must be a state of affairs that makes $p$ or $q$ true; this statement is false in the case in which both $p$ and $q$ are not verified. Consequently, a proposition with $n$ disjuncts $p_1\vee p_2\vee\dots\vee p_n$ is true if and only if at least one $p_i$ is true, and false otherwise. Hence, the logical operator $\vee$ retains its classical meaning. Contrary to QL, in Bohmian mechanics one does not have to introduce a new operator for the quantum disjunction, since the primitive ontology of BM is always exclusively located in one specific support of the quantum mechanical wave function, as already underlined in the previous section. This fact ensures that its evolution will evolve in only one of the macroscopically disjoint possible measurement outcomes. As Bacciagaluppi clearly underlines ``the configuration of the system is located only in one of these different components, and this is already a matter of classical logic. The cat is (classically) either alive or dead, because the particles that compose it are (classically) either in the live component or the dead component of the quantum state.'' (\cite{Bacciagaluppi:2009}, p.\ 72). Similarly, taking into account the previous example of the spin measurement along the $z$-axis, the particle will have spin up or spin down in the case in which it will be located above or under the symmetry line respectively. Hence, one can model this experimental situation with a classical disjunction. This marks a notable difference between BM and standard QM.
\vspace{2mm}

\item \textbf{Classicality of} $\neg$: Given a proposition $p$, $p$ is false in those cases in which it is not verified, i.e.\ in cases in which there is a physical state of affairs that makes $\neg p$ true. Thus, also the logical operator ``$\neg$'' maintains its classical meaning. For instance, the negations of the sentences ``the velocity of the particle $i$ is $v$'' and ``the position of the particle $i$ is $q$'' are expressed by the statements ``the velocity of the particle $i$ is not $v$'' and ``the position of the particle $i$ is not $q$'' respectively, having positions and velocity of this particle values different from $v$ and $q$.  In addition, let us consider how BM handles the case of two-valued operators such as spin observables---recalling that these do not represent genuine properties of Bohmian particles. In QL one models the possible outcomes of a spin measurement as two different and complementary tests, suppose $T$ for spin-up and $T^{\perp}$ for spin-down along a given axis, say $z$. Therefore, the negation of the proposition ``a certain system passed the test $T$ with probability 1'', corresponds to the following statement ``the system passed with probability 1 the test $T^{\perp}$, the orthogonal complement of the test $T$''. Such an interpretation can be retained in BM as well. 
In the context of this theory to state that a particle $i$ does not have $z-$spin-up is equivalent to assert that $i$ has not been found in a certain position in space corresponding to the location of the detector associated with a certain state of the measured operator. Hence, the negation of the proposition ``the particle $i$ has $z-$spin up'' is translated into the statement ``the particle $i$ is not at $q$'', where $q$ is the position where the particle would have been found if it had $z-$spin up. Furthermore, knowing that in spin measurements a particle will be necessarily found either in the state spin-up or spin-down, in BM one will add the information that the particle $i$ has been found in another position corresponding to the location of the detector associated with a the state ``spin-down''. Thus, one can recover the usual interpretation of negation as (ortho-)complementation as in the the logic of classical and quantum mechanics. As a consequence, negation has the following properties also in BM: (i) $\neg\neg p=p$, (ii) $p\wedge \neg p=0$ and (iii) $p \vee \neg p=1$. 
\end{itemize}

From this discussion it is possible to claim that the logical operators in BM retain their classical interpretation, and that complex propositions are formed with the usual logical connectives; hence, we can conclude that the logic of propositions concerning the Bohmian particles is classical. As a direct consequence, the distributivity law holds in such a theory. In order to discuss this fact, let us take into account the example already introduced in Section \ref{QL}. Considering the three propositions
\begin{itemize}
\item $p =$ the particle $i$ has $x-$spin-up;
\item $q =$ the particle $i$ has $y-$spin-up;
\item $r =$ the particle $i$ has $y-$spin-down,
\end{itemize}
\noindent we were able to show the failure of distributivity as follows. In the first place we assumed that the particle $i$ has $x-$spin-up, so that the proposition $p\wedge(q\vee r)$ expresses a true statement, saying that the particle has $x-$spin-up and simultaneously that $i$ has the spin along the $y$ axis either up or down---a condition which must be true in virtue of the quantum logical disjunction. However, given the non-commutativity between these spin operators, $[S_x, S_y]\neq 0$, it follows that the two conjunctions ``the particle $i$ has $x-$spin-up and $y-$spin-up'' and ``the particle $i$ has $x-$spin-up and $y-$spin-down'' are both false; hence, their disjunction false as well. Therefore, in standard QM the distributive law fails. 

On the contrary, this example does not constitute a violation of distributivity in BM. To show this simple fact, it is sufficient to recall that in the Bohmian context spin measurements reduce to measurements of particles' positions. Consequently, the sentence ``the particle $i$ has spin up along the $x$ axis'' can be translated into the following proposition:
\begin{quote}
``the particle $i$ has been found after a $x-$spin measurement at the position $q$ in the spatial region $\Delta$, where the detector associated with the state $x-$spin-up is located''. 
\end{quote}
\noindent Let's call this statement $p'$ and assume that $p'$ is true. Similarly, we can translate the sentences ``the particle $i$ has $y-$spin-up'' with
\begin{quote}
``the particle $i$ has been found after a $y-$spin measurement at the position $q'$ in the spatial region $\Delta'$, where the detector associated with the state $y-$spin-up is located'',
\end{quote}
\noindent and ``the particle $i$ has $y-$spin-down'' with 
\begin{quote}
``the particle has been found after a $y-$spin measurement at the position $q''$ in the region $\Delta''$, where the detector associated with the state $y-$spin-down is located''. 
\end{quote}
\noindent Let us call these statements $q'$ and $r'$ respectively. Clearly in these cases we require that these regions are mutually disjoint---$\Delta\cap\Delta'=\emptyset, \Delta\cap\Delta''=\emptyset, \Delta'\cap\Delta''=\emptyset$, and therefore $q\neq q' \neq q''$. 

Since the property of being spatially localized necessarily forbids that a particle is simultaneously in two places, and given that in BM we employ the classical disjunction, it follows that the l.h.s side of the distributivity law $p'\wedge(q'\vee r')$ expresses a \emph{false} statement, stating that the particle $i$ has position $q$ and simultaneously has position either $q'$ or $q''$, but since all these positions are mutually incompatible, it is not the case that a Bohmian particle can be simultaneously in more than one of such locations. Analogously, the r.h.s. of the distributivity law $(p'\wedge q')\vee(p'\wedge r')$ is false as well, given that it claims that the particle $i$ has simultaneously position $q$ and $q'$ or simultaneously position $q$ and $q''$, which is obviously impossible. Consequently, being both sides of the distributivity law false, the bi-conditional is true in virtue of the truth table of this logical operator. Therefore, the distributivity law is valid in this example, as showed by its truth table below:

\begin{center}
\includegraphics[scale=0.7]{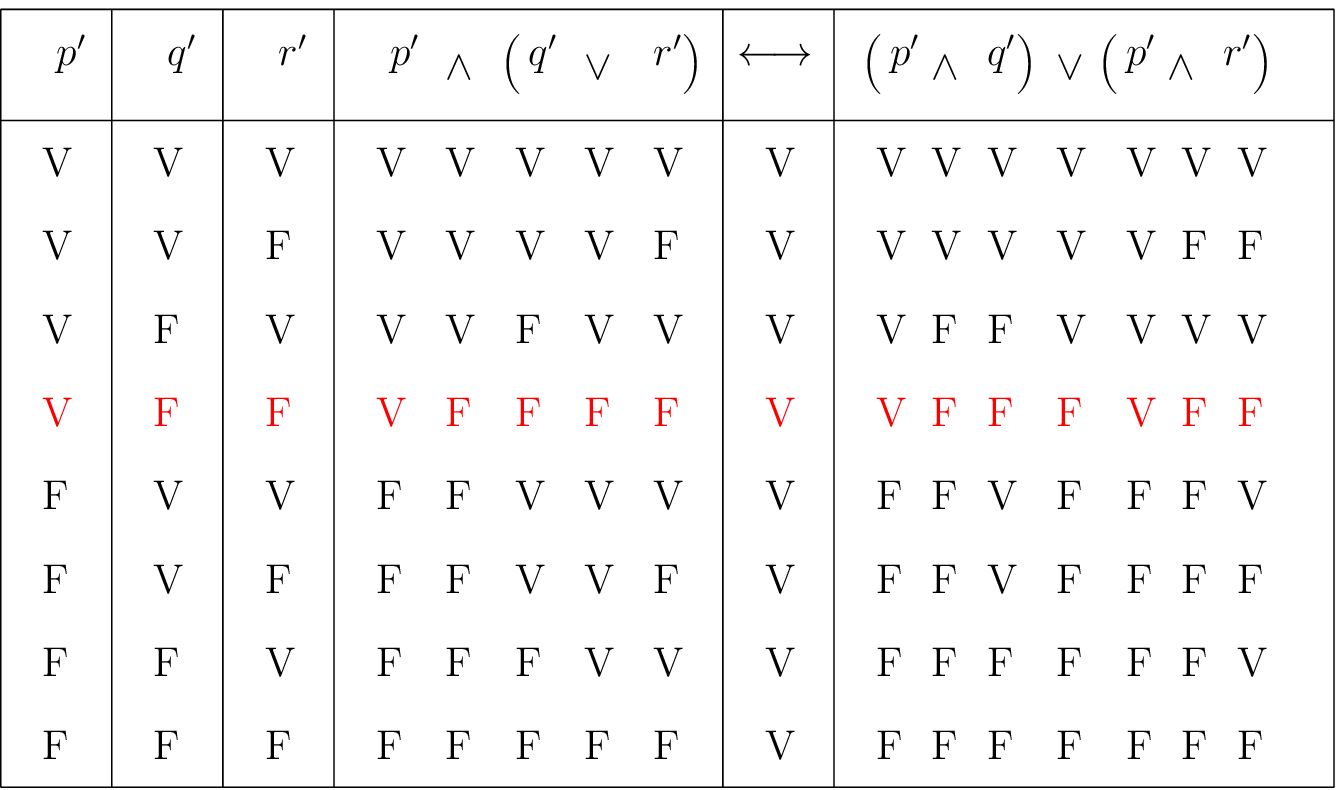}
\begin{quote}
\footnotesize{Fig. 5: Truth table of the distributivity law. The red line represents the truth value of the statements discussed in the above example.}
\end{quote}
\end{center}

Another example in which distributivity holds in BM is given by the following statements $p$ = ``the particle $i$ has location $q$ in space at a certain time $t$'' and $v_i$ = ``the particle $i$ has velocity $v_i$ at $t$'', where $i=1, \dots, n$ correspond to the different values the velocity can take. As already stressed, in BM velocity is a secondary property of the Bohmian corpuscles, since it is the first derivative of position over time, then, it is a property attributable to the particles. If we assume that $p$ is true, the statement $$p\wedge (v_1 \vee v_2, \vee \dots \vee v_n) = (p \wedge v_1) \vee (p \wedge v_2) \vee \dots \vee (p \wedge v_n)$$ will be true as well, since one of the disjuncts corresponds to the exact properties of the particle $i$ at a given time $t$, therefore both sides of the distributivity law are true also in this specific case, contrary to the case of standard quantum theory (in which such a statement would be meaningless!). 
\vspace{2mm}

In sum, we can conclude our analysis claiming that Bohmian mechanics ``is entirely classical at the level of kinematics (particles moving in space and time), and which is quantum only as regards its `new dynamics' (as in the title of de Broglie's paper). Thus, the way de Broglie-Bohm theory explains the effectiveness of classical logic at the macroscopic level is that it is already the logic that is operative at the hidden (`untestable') level of the particles'' (\cite{Bacciagaluppi:2009}, p.\ 71).
Concluding, in this section we showed in the first place that the Bohmian framework implements a classical logical structure at the level of its primitive ontology. This fact is sufficient to explain why the classical connectives maintain their meanings, without the need to introduce neither new quantum connectives, nor alternative interpretations of the logical operators. In the second place, we claimed that the distributive law is not violated in the context of Bohmian mechanics.

\section{Conclusion}
\label{conc}

In this essay we asked whether quantum physics necessarily implies a revision of the classical propositional calculus, or if it possible to restore a classical logical picture in the quantum domain. We answered in the positive to this question. Analyzing the primitive ontology of Bohmian mechanics and its theory of measurement, we provided a simple explanation of the reasons for which the propositions about Bohmian particles generate a classical logical structure. More precisely, we argued that statements about experiments and assertions regarding observables reduce to---i.e.\ are translatable into---propositions concerning the particle configuration of a certain experimental situation governed by the dynamical laws \eqref{SE}-\eqref{guide}. Building on this fact, we claimed that in BM the logical connectives retain their classical meanings. As a consequence, the distributivity law holds in this theoretical framework. 

Another interesting consequence of this result is that one can provide a justification for the classical logical structure which is generally valid in the macroscopic regime. Given that (i) in a Bohmian universe the macroscopic world is ontologically reduced to---thus, metaphysically dependent upon---the configurations of particles' and their deterministic dynamics, and (ii) these generate a classical propositional calculus, it follows that such a logic will be maintained at the macroscopic regime. However, we invite the reader to consider the validity of the present argument with grain of salt, since we did not enter in discussion about the empirical nature of logic.

In sum, from the discussion contained in Sections \ref{BM} and \ref{CL}, we can conclude that the classical logic underlying BM should be interpreted as a further consequence of the theory's primitive ontology. Thus, Bohmian mechanics constitutes a counterexample to those claims for which quantum physics \emph{necessarily} implies a revision of logic. This notable feature of the PO of this theory unfortunately has not been emphasized in the literature, and we hope to have, at least partially, filled this lacuna. Moreover, it would be an interesting future work to try to extend the present analysis of BM in order to understand the logical structure of other well-known PO theories such as Bohm's original pilot-wave theory or the Ghirardi-Rimini-Weber theory implementing a flash and matter density ontology. In particular, it is less trivial to establish what sort of logical structure underlies Bohm's pilot-wave theory and Ghirardi-Rimini-Weber with matter density ontology, since both frameworks admit the possibility to have superpositions of the $\psi$-field on the one hand, and the matter density field on the other, both entities defined in physical space. In this manner, it will be possible to investigate in more detail whether the rehabilitation of classical logic in quantum context is a general consequence of all theories implementing a clear primitive ontology.

\clearpage
\bibliographystyle{apalike}
\bibliography{PhDthesis}
\end{document}